\documentclass[aps,prd,twocolumn,preprintnumbers,showpacs,nofootinbib]{revtex4-1}
\usepackage{epsfig}
\usepackage[colorlinks,linkcolor=blue,anchorcolor=blue,citecolor=blue,urlcolor=blue,breaklinks=true]{hyperref}
\usepackage{graphics}
\usepackage{color}
\usepackage{graphicx}
\usepackage{amsmath}
\usepackage{amsfonts}
\usepackage{upgreek}
\usepackage{slashed}
\usepackage{color}
\usepackage[group-separator={,}]{siunitx}

\newcommand{\xg}{\langle x \rangle_g}

\begin{document}

\preprint{MSUHEP-18-014}

\title{\vspace{1.0in} {\bf Gluon Quasi-PDF From Lattice QCD}}

\author{Zhou-You Fan$^{1}$, Yi-Bo Yang$^{1,2}$, Adam Anthony$^{1}$,  Huey-Wen Lin$^{1,3}$, and Keh-Fei Liu$^{4}$
}
\affiliation{
$^{1}$\mbox{Department of Physics and Astronomy, Michigan State University, East Lansing, MI 48824, USA}\\
$^{2}$\mbox{Institute of Theoretical Physics, Chinese Academy of Sciences, Beijing 100190, China}\\
$^{3}$\mbox{Department of Computational Mathematics,  Science and Engineering, Michigan State University, East Lansing, MI 48824, USA}\\
$^{4}$\mbox{Department of Physics and Astronomy, University of Kentucky, Lexington, KY 40506, USA}\\
}

\begin{abstract}
We present the first attempt to access the $x$-dependence of the gluon unpolarized parton distribution function (PDF), based on lattice simulations using the large-momentum effective theory (LaMET) approach. The lattice calculation is carried out with pion masses of 340 and 678~MeV on a 2+1-flavor DWF configuration with lattice spacing $a=0.111$~fm, for the gluon quasi-PDF matrix element with the nucleon momentum up to 0.93 GeV. 
Taking the normalization from similar matrix elements in the rest frame of the nucleon and pion, our results for these matrix elements are consistent with the Fourier transform of the global fit  CT14 and PDF4LHC15 NNLO of the gluon PDF, within statistical uncertainty and the systematic one up to power corrections, perturbative ${\cal O}(\alpha_s)$ matching and the mixing from the quark PDFs.
\end{abstract}


\maketitle

\textit{Introduction: }The unpolarized parton distribution function (PDF) is the probability density for finding the corresponding parton with a certain longitudinal momentum fraction $x$ {in an infinite-momentum hadron} at $\overline{\text{MS}}$ renormalization scale $\mu$, {that satisfies the hadron momentum sum rules,
\begin{equation}
\int^1_0 \text{d}x\ x\ \left(g(x,\mu)+\sum_{q=u,\bar{u},d...} q(x,\mu)\right) = 1,
\end{equation}
where $g(x)$ and $q(x)$ are the unpolarized gluon and quark PDFs respectively. In the leading-twist collinear factorization, PDFs are process-independent and encode the intrinsic information of the quark and gluon inside the hadron. Even though the quark and glue momentum fractions are roughly half and half, their PDFs are quite different and the constraint from a given process can be differ by an order of magnitude or more.}

{For example, }although $g(x)$ contributes at next-to-leading order to the deep inelastic scattering (DIS) cross section {, where $q(x)$'s dominates}, it enters at leading order in jet production. Top-quark pair production at the LHC can provide significant constraints to the global fit of $g(x)$ at $x>0.1$ region~\cite{Czakon:2013tha}, and small-$x$ ($x<10^{-4}$) region of $g(x)$ is strongly constrained by charm production at high energies~\cite{Gauld:2015yia}. Thus, many fits have been done to constrain $g(x)$ by combining data from DIS and jet-production cross sections. {It is the phenomenologcial approach to determine PDFs: }With more experimental data and better fit approaches, our understanding of PDFs from experiments continues to improve.

{The theoretical approach, which is independent of the experiments and their fits, targets to extract PDFs from the first principle calculation of QCD.}
On the theoretical side, the unpolarized gluon PDF is defined by the Fourier transform (FT) of the lightcone correlation in the hadron,
\begin{align}\label{eq:gluon_PDF}
 g(x,\mu)&=\int\frac{\text{d}\xi^-}{\pi x}e^{-ix\xi^-P^+}\nonumber\\
           &\quad\quad \left\langle P|\big[F^+_{\ \mu}(\xi^-)U(\xi^{-},0)F^{\mu +}(0)\big](\mu)|P\right\rangle,
\end{align}
where $ \xi^{\pm}=\frac{1}{2}(\xi^0\pm\xi^3)$ is the spacetime coordinate along the lightcone direction, the hadron momentum $P_{\mu}=(P_0,0,0,P_z)$, $|P\rangle$ is the hadron state with momentum $P$ with the normalization $\langle P|P\rangle=1$, $\mu$ is the $\overline{\text{MS}}$ renormalization scale of the glue operator, $U(\xi^{-},0)=\mathcal{P}\exp(-ig\int^{\xi^-}_0\text{d}\eta^-A^+(\eta^-))$ is the lightcone Wilson link from $\xi^{-}$ to 0 with $A^+$ being the gauge potential in the adjoint representation, and $F_{\mu\nu}=T^aG^a_{\mu\nu}=T^a(\partial_{\mu}A^a_{\nu}-\partial_{\nu}A^a_{\mu}-gf^{abc}A^b_{\mu}A^c_{\nu})$ is the gauge field tensor. Based on such a definition, all the odd moments vanish due to the parity of the glue matrix elements, while the even ones survive. 

Even though the definition in Eq.~\ref{eq:gluon_PDF} involves a Minkowski spacetime correlation and is infeasible to construct in a Euclidean lattice simulation, its second moment is calculable in Euclidean space as the matrix elements of local operators:
\begin{align}
\xg&\equiv \int_0^1 x\ g(x) \text{d}x=\frac{1}{P^+}\langle P|F^+_{\mu}(0)F^{\mu +}(0)|P\rangle\label{eq:def_0}\\
   &=\frac{1}{P_z}\langle P|\overline{T}^{tz}(0)|P\rangle\nonumber\\
   &=\frac{P_0\langle P|\overline{T}^{zz}(0)|P\rangle}{\frac{1}{4}P_0^2+\frac{3}{4}P_z^2}=\frac{P_0\langle P|\overline{T}^{tt}(0)|P\rangle}{\frac{3}{4}P_0^2+\frac{1}{4}P_z^2},\label{eq:def_1}
\end{align}
where the gauge energy-momentum tensor $\overline{T}^{\mu\nu}=F^{\mu}_{\ \rho}F^{\rho\nu}-\frac{1}{4}g^{\mu\nu}F^{\tau}_{\ \rho}F^{\rho}_{\ \tau}$. Note that all the definitions in Eq.~(\ref{eq:def_1}) are frame-independent and can be calculated in a frame far from the infinite momentum. Moreover, the latter two definitions can be used to carry out the calculation in the rest frame of the hadron. Lattice calculations of $\xg$ in the nucleon~\cite{Horsley:2012pz,Deka:2013zha,Alexandrou:2016ekb,Alexandrou:2017oeh,Yang:2018bft,Yang:2018nqn} have been significantly refined in the last decade, while calculations of moments beyond the second moment are still absent. 

Based on the large-momentum effective theory (LaMET)~\cite{Ji:2013dva,Ji:2014gla} approach, a proper definition of the gluon quasi-PDF inspired by the last right-hand side of Eq.~(\ref{eq:def_1}) is
\begin{equation}
\tilde{g}(x,P_z^2,\mu) = \int \frac{\text{d}z}{\pi x} e^{-ix zP_z} \tilde{H}^R_0(z,P_z,\mu),
\end{equation}
where $\tilde{H}^R_0(z,P_z,\mu)$ is the gluon quasi-PDF matrix element
\begin{align}
\tilde{H}_0(z,P_z)&=\langle P|{\cal O}_0(z)|P\rangle,\\
{\cal O}_0 &\equiv
  \frac{P_0\left({\cal O}(F^t_{\ \mu}, F^{\mu t};z) - \frac{1}{4}g^{tt}{\cal O}(F^{\mu}_{\ \nu}, F^{\nu}_{\ \mu};z)\right)}
       {\frac{3}{4}P_0^2 + \frac{1}{4}P_z^2},\nonumber
\end{align}
 renormalized at the scale $\mu$ with ${\cal O}(F^{\rho}_{\ \mu}, F^{\mu \tau};z)=F^{\rho}_{\ \mu}(z)U(z,0)F^{\mu \tau}(0)$. When $z=0$, $\tilde{H}_0(0,P_z)$ is a local operator and equals to $\xg$. In the large momentum limit, only the leading twist contribution in $\tilde{g}(x)$ survives, and then $\tilde{g}(x)$ can be factorized into the the gluon PDF $g(y)$ and a perturbative calculable kernel ${\cal C}(x,y)$, up to mixing with the quark PDF and the higher-twist corrections ${\cal O}(1/P_z^2)$. 
 
{Since the Lattice calculation of $\tilde{H}_0(z,P_z)$ is under the lattice regularization, a non-perturbative renormalization (NPR) of the glue operators ${\cal O}_0(z)$ is required to convert $\tilde{H}_0(z,P_z)$ into that under the $\overline{\text{MS}}$ scheme with the perturbative matching in the continuum. This can be achieved following the glue NPR strategy introduced in Ref.~\cite{Yang:2018bft} just recently for $\xg$.}
 

{As shown in Refs. \cite{Zhang:2018diq,Li:2018tpe}, the ${\cal O}(F^{z}_{\ \mu}, F^{\mu z};z)$ and ${\cal O}(F^{\mu}_{\ \nu}, F^{\nu \mu};z)$ ($\mu, \nu\neq z$) structures in ${\cal O}_0$ should be renormalized separately before combined together, but its linear divergence~\cite{Wang:2017qyg,Wang:2017eel} is an overall multiplicative factor depending on the Wilson-link length $z$. For the linear divergence introduced by the Wilson link, an empirical observation in the quark unpolarized quasi-PDF case is that, the non-perturbative RI/MOM renormalization constant with $p_z^R=0$ can be approximated by the nucleon iso-vector matrix element with $P_z=0$ in the $z<0.5$ fm region, with $\sim$10\% deviation, while the systematic uncertainties due to the hadron IR structure is hard to estimate~\cite{Liu:2018uuj}. If the gluon case is similar, the linear divergence of the gluon quasi-PDF matrix element can be removed by defining the ``ratio renormalization" (similar to the reduce Ioffe-time distribution considered in the quark case~\cite{Radyushkin:2017cyf,Orginos:2017kos,Broniowski:2017gfp})
\begin{align}
\tilde{H}_0^{Ra}(z,P_z,\mu)=\frac{\tilde{H}_0^{\overline{\text{MS}}}(0, 0, \mu)}{\tilde{H}_0(z, 0)}\tilde{H}_0(z,P_z)
\end{align}
as an approximation of the RI/MOM renormalized one, with $\tilde{H}_0^{Ra}(z,P_z,\mu)=\xg^{\overline{\text{MS}}}(\mu)$.
}

After the renormalization, both the quark and gluon PDF contribute to the factorization of the gluon qausi-PDF~\cite{Wang:2017qyg}, and the case with the gluon quasi-PDF operator defined here will be investigated in a future study. In this work, we will calculate the gluon quasi-PDF matrix element and apply the ``ratio renormalization" to have a glimpse on the range of $z$ and $P_z$ one can reach on the lattice, and compare it with the FT of the gluon PDF.


\textit{Numerical setup: }The lattice calculation is carried out with valence overlap fermions on 203 configurations of the $2+1$-flavor domain-wall fermion gauge ensemble ``24I''~\cite{Aoki:2010dy} with $L^3\times T=24^3\times 64$, $a=0.1105(3)$~fm, and $M_\pi^\text{sea}$=330~MeV. 
For the nucleon two-point function, we calculate with the overlap fermion and loop over all timeslices with a 2-2-2 $Z_3$ grid source and low-mode substitution~\cite{grid1,grid2}, and set the valence-quark mass to be roughly the same as the sea and strange-quark masses (the corresponding pion masses are 340 and 678~MeV, respectively). Counting independent smeared-point sources, the statistics of the two-point functions are $203\times64\times8\times2=\num{207872}$, where the last factor of 2 coming from the averaging between the forward and backward nucleon propagators.


On the lattice, ${\cal O}_0$ is defined by
\begin{align}
{\cal O}_0=-\frac{P_0\left({\cal O}_E(F_{t\mu},F_{\mu t},z) - \frac{1}{4}{\cal O}_E(F_{\mu\nu}, F_{\nu\mu};z)\right)}
       {\frac{3}{4}P_0^2 + \frac{1}{4}P_z^2}
\end{align}
where \mbox{${\cal O}_E(F_{\rho\mu},F_{\mu \tau},z)=2\textrm{Tr}\big[F_{\rho\mu}(z)U(z,0)F_{\mu \tau}(0)U(0,z)\big]$} is defined in the Euclidean space with the gauge link $U(z,0)$ in the fundamental representation, and  
 the clover definition of the field tensor $F_{\mu\nu}$ is the same as that used in our previous calculation of the glue momentum fraction~\cite{Yang:2018bft}.
 
The choice for the quasi-PDF operator is not unique. Any operator that approaches the lightcone one in the large-momentum limit is a candidate, such as the other choices inspired by Eq.~(\ref{eq:def_1})
\begin{align}
{\cal O}_1(z) &\equiv \frac{1}{P_z}{\cal O}(F_{t\mu},F_{z\mu};z),\nonumber\\
{\cal O}_2(z) &\equiv
  \frac{P_0\left({\cal O}(F_{z\mu}, F_{\mu z};z)-\frac{1}{4}g^{zz}{\cal O}(F_{\mu\nu}, F_{\nu\mu};z)\right)}
       {\frac{1}{4}P_0^2 + \frac{3}{4}P_z^2},
\end{align}
as well as 
\begin{equation}
{\cal O}_3(z) \equiv \frac{1}{P_0}{\cal O}(F_{z\mu},F_{\mu z};z)
\end{equation}
proposed in Ref.~\cite{Ji:2013dva}. These alternative operators ${\cal O}_{1,2,3}$ can be defined on the lattice similarly.  As we will address in the latter part of this work, the quasi-PDF using ${\cal O}_{1,2,3}$ has larger higher-twist corrections and/or statistical uncertainty compared to that from using ${\cal O}_{0}$.
 
The bare glue matrix element $\tilde H_0(z, P_z)$ with the Wilson link length $z$ and nucleon momentum $\{0,0,P_z\}$ can be obtained from the derivative of the summed ratio following the recent high-precision calculation of nucleon matrix elements~\cite{Berkowitz:2017gql,Chang:2018uxx},

\begin{figure}[htbp]
  \centering
  \includegraphics[width=0.5\textwidth]{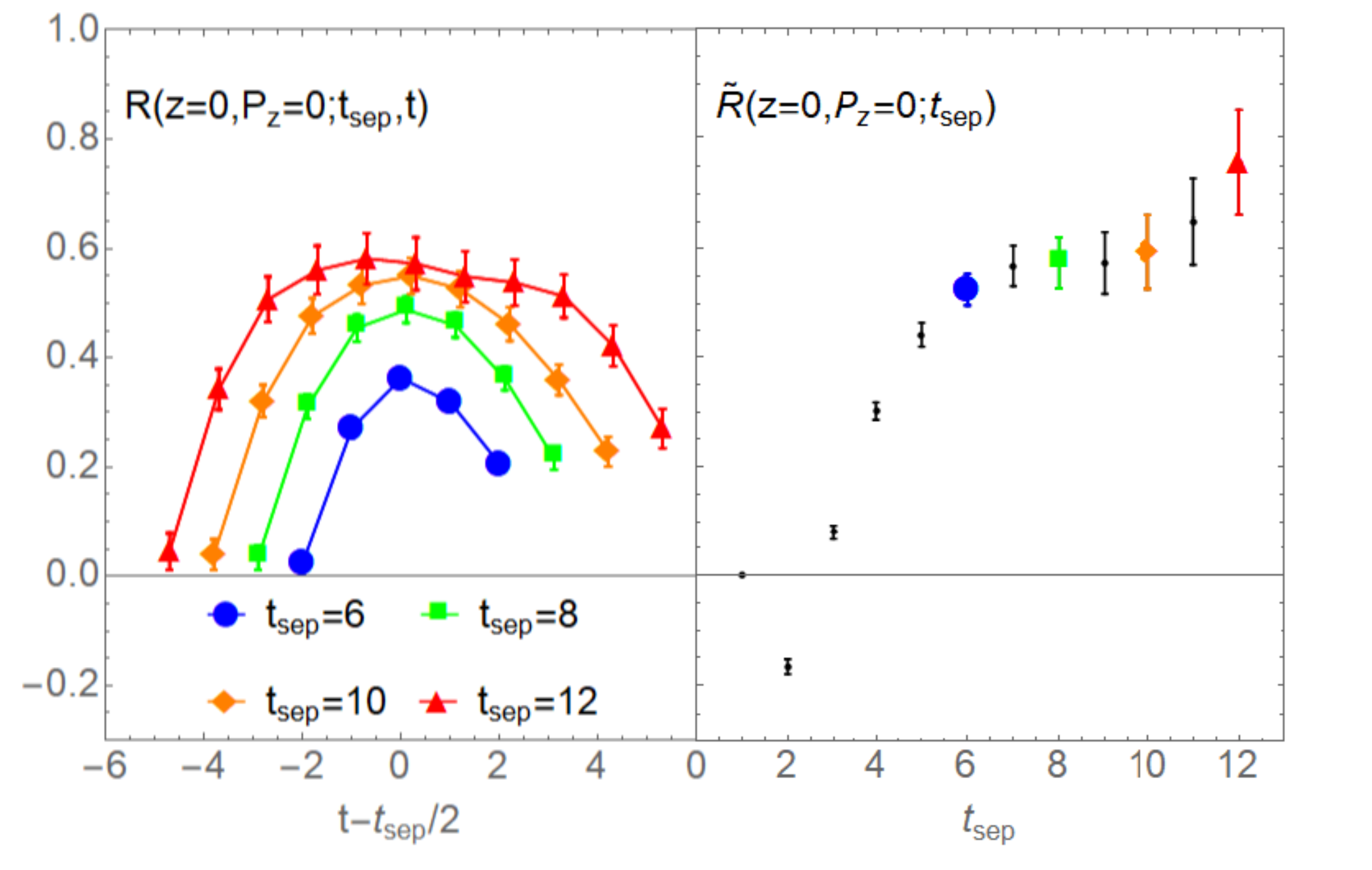}\\
  \caption{
  The ratio $R(t_{\rm sep},t)$ for $\tilde H_0(0,0)$ at different $t_{\rm sep}$ as a function of operator insertion time $t$ (left panel), and the ratio $\tilde{R}(t_{\rm sep})$  as a function of source-sink seperation $t_{\rm sep}$ (right panel).  Four colored points in the right panel corresponds to the $\tilde{R}$ at the separations plotted in the left-panel. 
   }\label{fig:ratio}
\end{figure}

\begin{align}
\tilde{R}(z,P_z;t_{\rm sep})&=\sum_{0<t<t_{\rm sep}}R(z,P_z;t_{\rm sep},t) \nonumber\\
 &-\sum_{0<t<t_{\rm sep}-1}R(z,P_z;t_{\rm sep}-1,t) \nonumber\\
 &=\tilde H_0(z,P_z)+\mathcal{O}(e^{\Delta m t_{\rm sep}}),
\end{align}
where
\begin{multline}
 R(z,P_z;t_{\rm sep},t)\equiv \nonumber\\
 \frac{E\langle 0|\Gamma^e\int \text{d}^3y e^{-iy\cdot P}\chi(\vec y,t_{\rm sep}){\cal O}_0(z;t)\chi(\vec 0,0)|0\rangle}{(\frac{3}{4}E^2+\frac{1}{4}P_z^2)\langle 0|\Gamma^e\int \text{d}^3y e^{-iy_3P_3}\chi(\vec y,t_{\rm sep})\chi(\vec 0,0)|0\rangle}
\end{multline}
and $\Gamma^e=\frac{1}{2}(1+\gamma_4)$. To further improve the signal of $\tilde{H}_0$, we applied up to 5 steps of HYP smearing on the glue operators.


\begin{figure}[htbp]
  \includegraphics[width=0.45\textwidth]{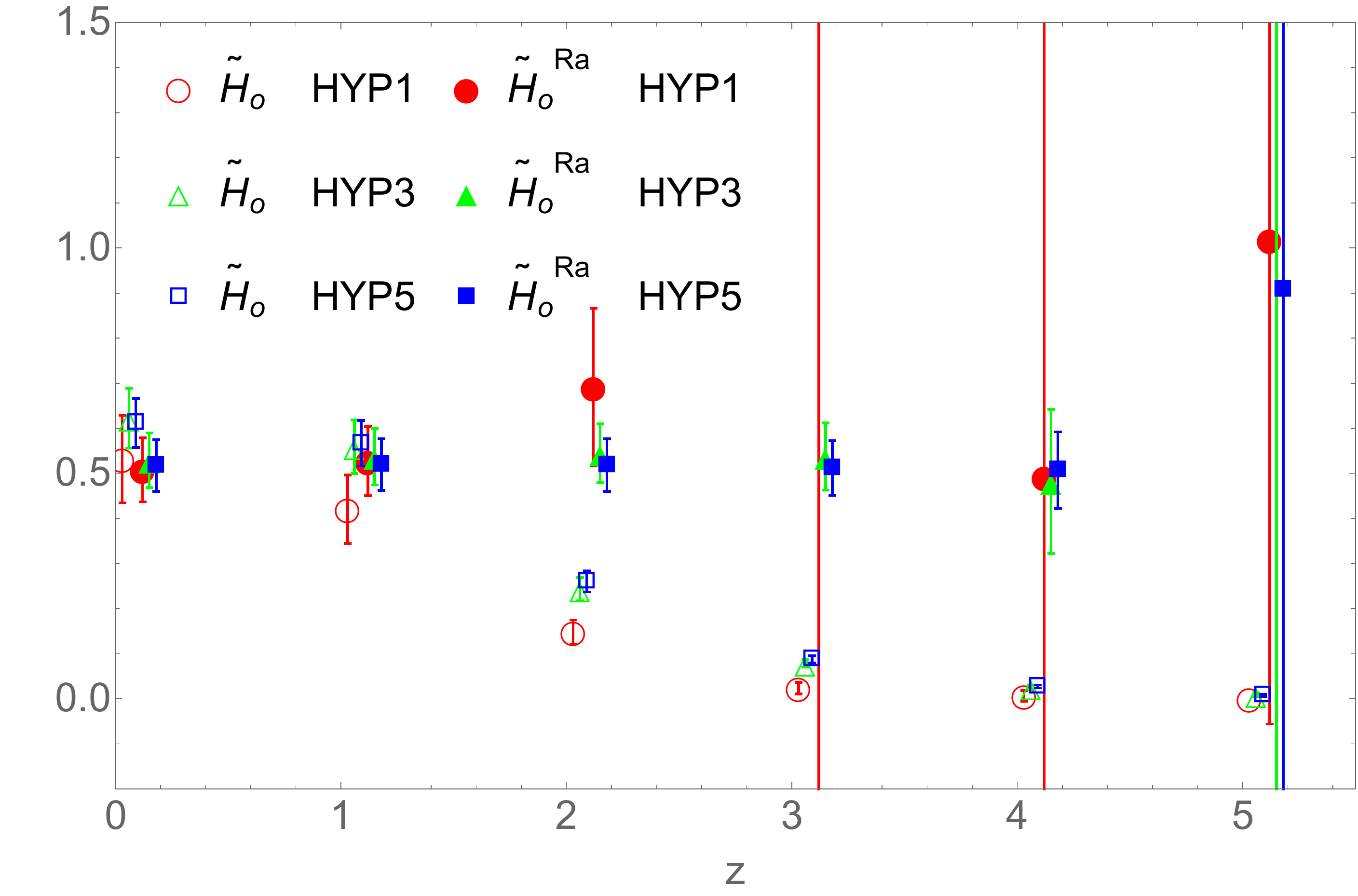}
  \caption{The bare $\tilde{H}(z,P_z=0.46\text{ GeV})$ and the renormalized one $\tilde{H}^{Ra}$ at 2~GeV with 1,3,5 HYP smearing steps, as functions of $z$. In $\tilde{H}^{Ra}$, the exponential falloff in the bare $\tilde{H}$ due to the linear divergence is obviously removed by the ``ratio renormalization factor" $Z(\mu,z)\equiv H^{\overline{\text{MS}}}_{0}(0,0,\mu)/\tilde{H}_0(z,0)$. Some data using the same HYP smearing steps are shifted horizontally to enhance the legibility. }\label{fig:z_dep}
\end{figure}

\textit{Results: }As illustrated in Fig.~\ref{fig:ratio} for $\tilde H_0(0,0)$ with 5 HYP smearing steps, the value of $\tilde{R}$ saturates after $t_{\rm sep}>6$ and a constant fit can provide the same result as what can be obtained from the two-state fit of $R$ with larger $t_{\rm sep}$. In the $t_{\rm sep}\gg t\gg 0$ limit, both $\tilde{R}$ and $R$ saturate to the same $\tilde H_0(0,0)=\xg^\text{bare}=0.55(8)$ as
  in the figure, while such a limit can be reached with smaller $t_{\rm sep}$ in the $\tilde{R}$ case.
Using the renormalization constant of $\xg$ in $\overline{\text{MS}}$ at 2~GeV with 5 steps of the HYP smearing calculated in Ref.~\cite{Yang:2018bft} of 0.90(10) and ignoring mixing from the quark momentum fraction, the $\overline{\text{MS}}$ renormalized $\xg^{\overline{\text{MS}}}(2\text{ GeV})=\tilde H^{Ra}_0(0,0,2\text{ GeV})=0.50(7)(5)$ agrees with the phenomenological determination 0.42(2)~\cite{Dulat:2015mca} within uncertainties.

\begin{figure}[htbp]
  \includegraphics[width=0.45\textwidth]{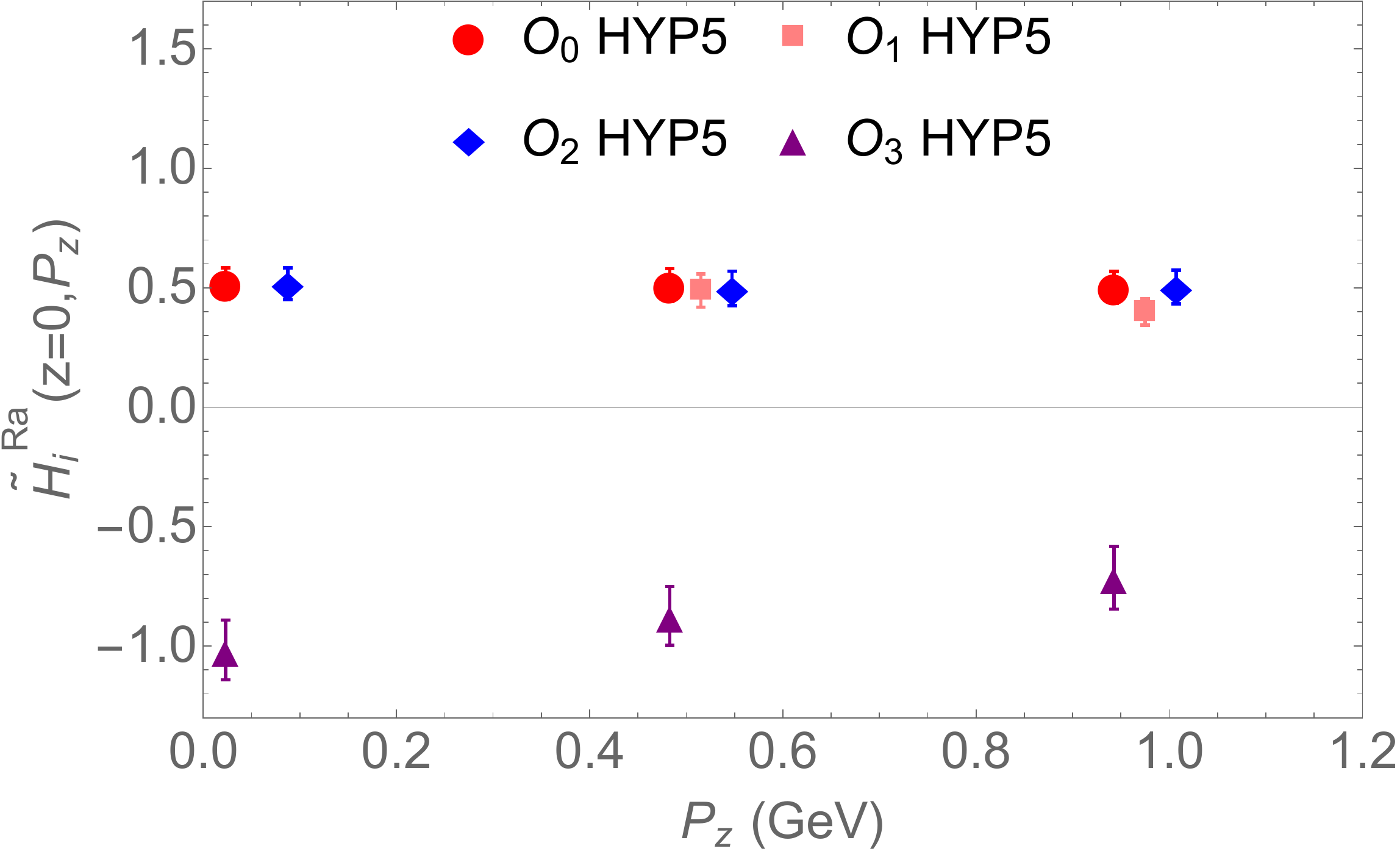}\\
  \includegraphics[width=0.45\textwidth]{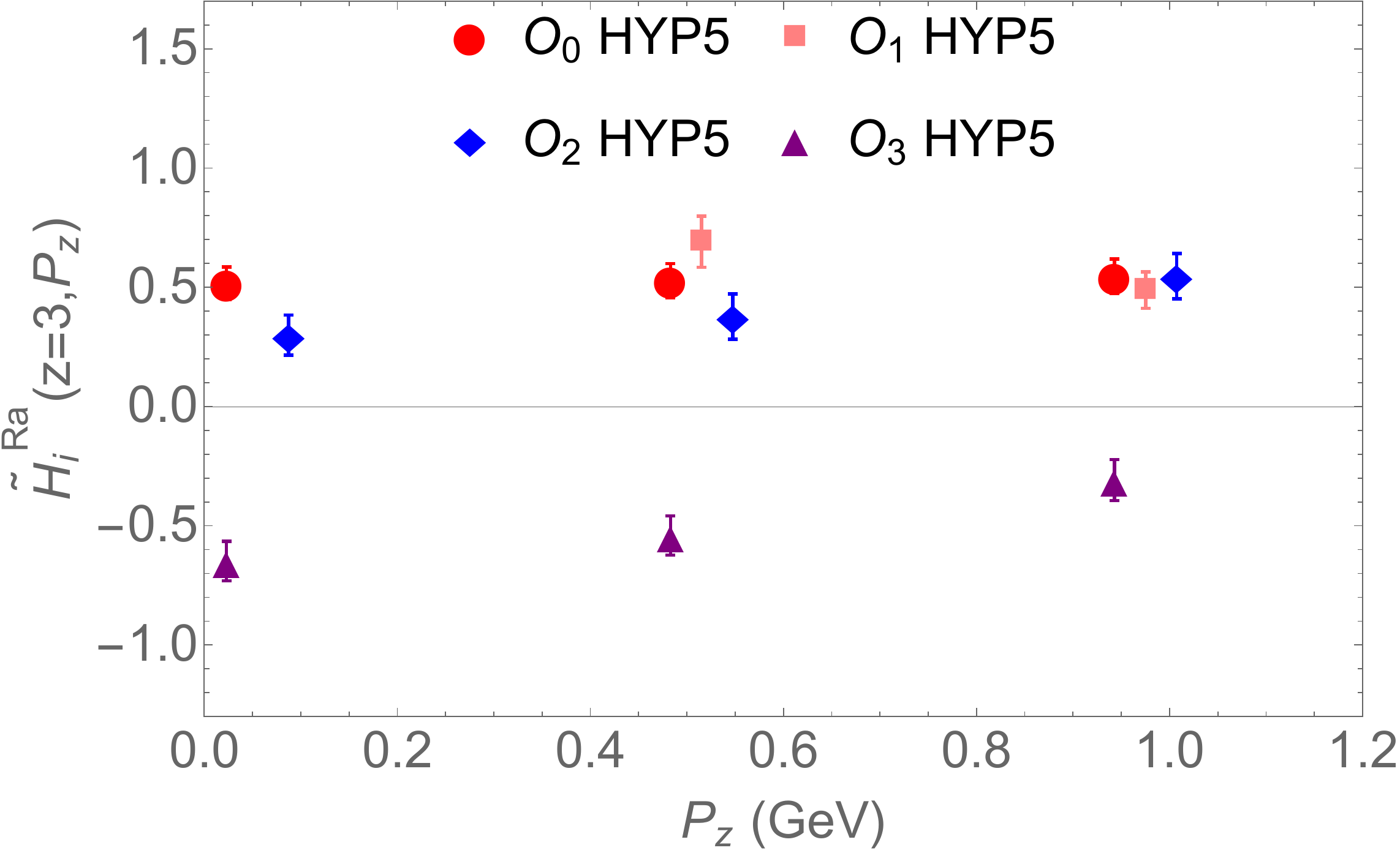}
  \caption{The renormalized $\tilde{H}^{Ra}_{i=0,1,2,3}(z,P_z)$ as a functions of $P_z$ at $z$=0 (top) and 3 (bottom). Some data with the same $P_z$ are shifted horizontally to enhance the legibility.  The case with ${\cal O}_{i=3}$ suffers from a large contamination from higher-twist distributions, while the results with ${\cal O}_{i=0,1,2}$ are consistent with each other, especially at larger $P_z$.}\label{fig:H_r}
\end{figure}

Due to its linear divergence~\cite{Wang:2017eel}, the bare $\tilde H_0(z,P_z)$ decays exponentially as $|z|$ increases. Fig.~\ref{fig:z_dep} shows the $z$ dependence of $\tilde H_0(z,P_z)$ with $P_z=0.46$~GeV and 1, 3 and 5 HYP smearing steps. It is obvious to see that the decay rates decreases when more steps of smearing are applied, since the corresponding linear divergence becomes smaller. Note that $\tilde H_0(z,P_z)$ is purely real and symmetric with respect to $z$; thus, we just plot the real part in the positive-$z$ region. The ``ratio renormalized" matrix elements $\tilde H^{Ra}_0(z,P_z)$ with different HYP smearing steps are consistent with each other, as shown in Fig.~\ref{fig:z_dep}, while more HYP smearing can reduce the statistical uncertainties significantly.

Then, we plot the ``ratio renormalized" $\tilde{H}^{Ra}_{i=0,1,2,3}(z=0,P_z)$ using $Z(\mu, z)\equiv \frac{\tilde{H}_0^{\overline{\text{MS}}}(0, 0, \mu)}{\tilde{H}_0(0, 0, \mu)}$ for the glue operator ${\cal O}_{i}$ with 5 HYP smearing steps and $P_z= 0.0$, 0.46, 0.92~GeV in the top panel of Fig.~\ref{fig:H_r}. All the cases with ${\cal O}_{i=0,1,2}$ provide consistent results, except ${\cal O}_{3}$ which suffers from large mixing with the higher-twist operator ${\cal O}(F^{\mu}_{\nu}, F^{\nu}_{\mu};z)$. With larger $P_z$, the value of $\tilde{H}^{Ra}_3(0, P_z)$ becomes less negative as higher-twist contamination becomes smaller.

The lower panel of Fig.~\ref{fig:H_r} shows $\tilde{H}^{Ra}_{i=0,1,2,3}(z=3,P_z)$ with different operators and  $P_z= 0.0$, 0.46, 0.92~GeV. The ${\cal O}_{3}$ case also suffers from large higher-twist contamination like the $z=0$ case; the results with ${\cal O}_{i=0,1,2}$ seem to be slightly different from each other at $P_z=0.46$~GeV, while the consistency at $P_z=0.92$~GeV is much better. Since the operators $O_{0,1,2}$ can provide consistent results but the uncertainty using $O_{0}$ is slightly smaller than the other two cases, we will concentrate on this case in the following discussion.

\begin{figure}[htbp]
  \includegraphics[width=0.45\textwidth]{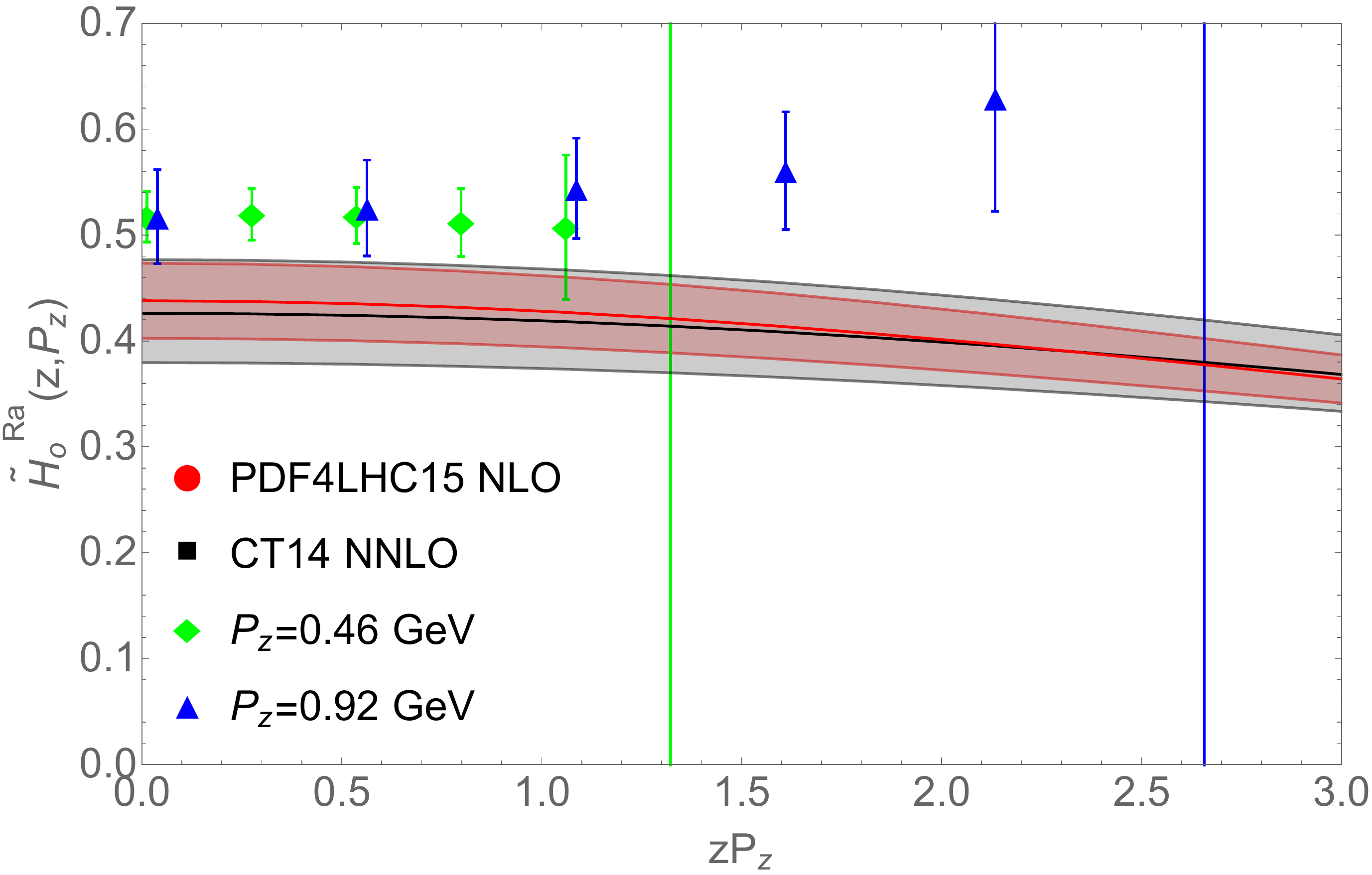}
  \includegraphics[width=0.45\textwidth]{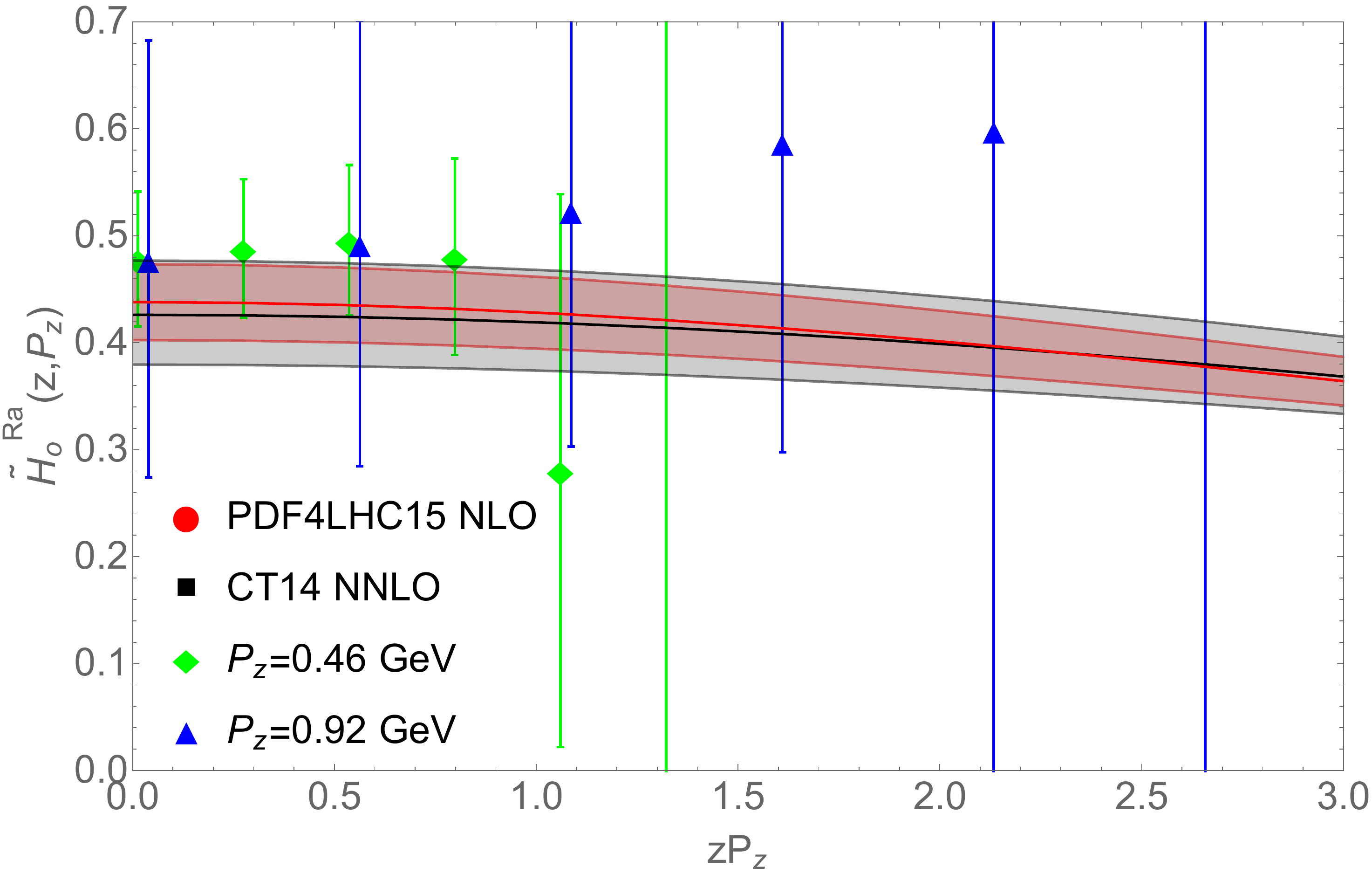}
  \caption{The final results of $\tilde{H}^{Ra}_0(z,P_z)$ at 678~MeV (top) and 340~MeV (bottom) pion mass as a functions of $zP_z$, in comparison with the FT of the gluon PDF from the global fits CT14~\cite{Dulat:2015mca} and PDF4LHC15 NNLO~\cite{Butterworth:2015oua}. The data with $P_z=0.92$~GeV are shifted horizontally to enhance the legibility. They are consistent with each other within the uncertainty.}\label{fig:final}
\end{figure}

\begin{figure}[htbp]
  \includegraphics[width=0.45\textwidth]{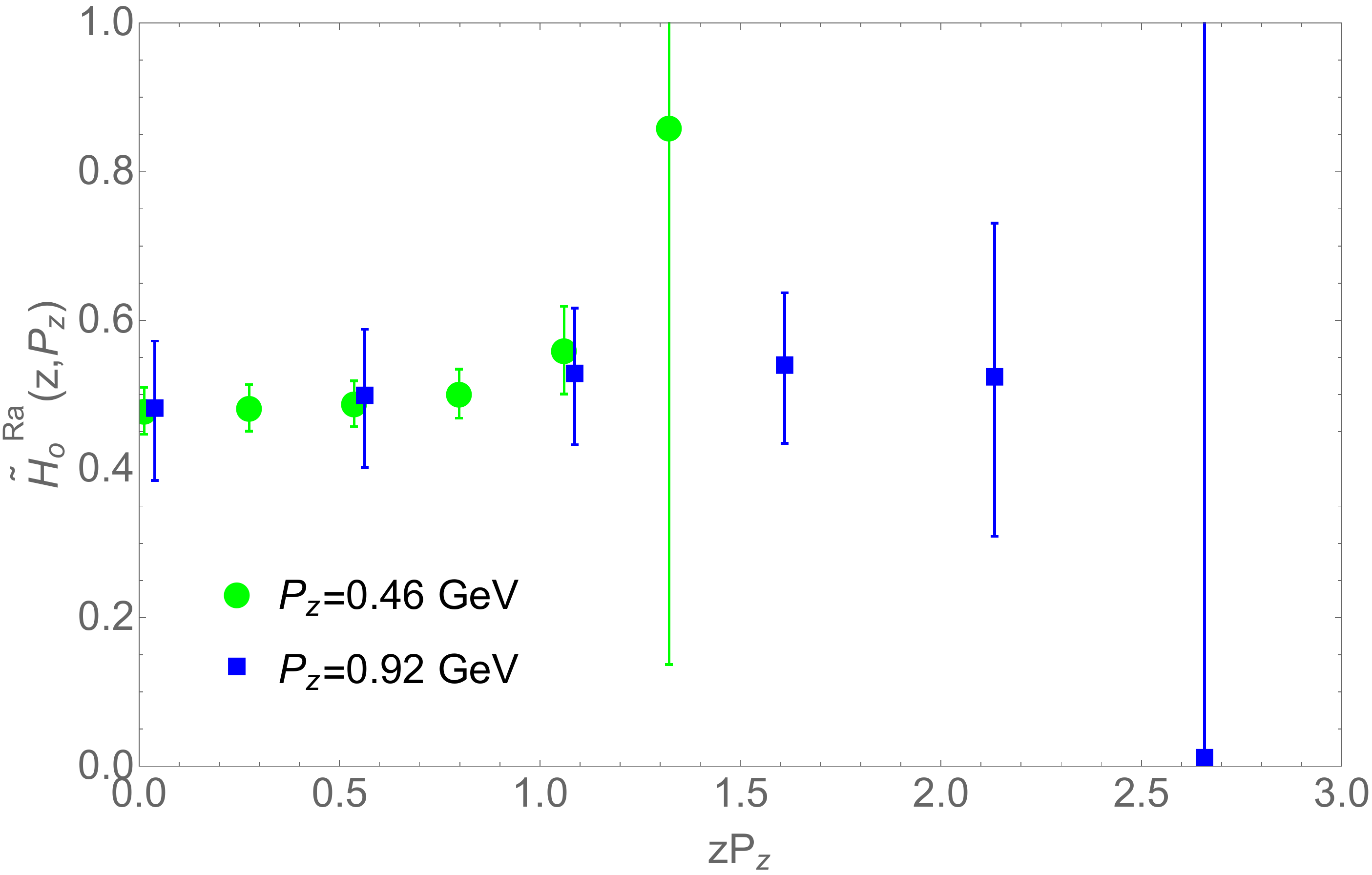}
  \caption{The similar figure for the pion gluon quasi-PDF matrix elements with $M_{\pi}=678$~MeV. The shape is quite similar to the case in Fig.~\ref{fig:final}.}\label{fig:final2}
\end{figure}

Finally, the coordinate-space gluon quasi-PDF matrix element ratios $\tilde{H}^{Ra}_{0}(z,P_z)$ are plotted in Fig.~\ref{fig:final}, compared with the corresponding FT of the gluon PDF, H(z, $\mu$=2~\textrm{GeV}), based on the global fits from CT14~\cite{Dulat:2015mca} and PDF4LHC15 NNLO~\cite{Butterworth:2015oua}. Since the uncertainties increases exponentially at larger $z$, our present lattice data with good signals are limited to the range $zP_z<$2 or so, and the values at different $zP_z$ are consistent with each other. At the same time, $H(z, 2~\textrm{GeV})$ doesn't changes much either in this region as in Fig.~\ref{fig:final}, as investigated in Ref.~\cite{Broniowski:2017gfp}. Up to perturbative matching and power correction at ${\cal O}(1/P_z^2)$, they should be the same, and our simulation results are within the statistical uncertainty at large $z$. The results at the lighter pion mass (at the unitary point) of 340~MeV is also shown in Fig.~\ref{fig:final}, which is consistent with those from the strange quark mass case but with larger uncertainties. We also study the pion gluon quasi-PDF (see Fig.~\ref{fig:final2}) and similar features are observed.

{In a recent work \cite{Yang:2018nqn} involving part of the present authors, the glue momentum fraction $\langle x\rangle^{\overline{\textrm{MS}}}$ (corresponds to $\tilde{H}^{Ra}(0)$ here) is calculated on configurations with different lattice spacing, valence and sea quark masses. The value of $\langle x\rangle^{\overline{\textrm{MS}}}$ tend to be slightly larger with smaller quark mass, but the dependence is weak. Thus it hints that the entire gluon distribution may be also insensitive to either the valence or sea quark mass given the current statistical errors, up to $\sim$ 400 MeV pion mass or so. The quark case is similar; thus we don't expect the gluon quasi-PDF and the mixing with the quark PDF through the factorization to be very sensitive to the quark mass unless the statistical uncertainty can be reduced significantly.}

If $\tilde{H}^{Ra}_{0}(z,P_z)$ keeps flat outside the region where we have good signal, the gluon quasi-PDF $\tilde{g}(x)$ will be a delta function at $x=0$ through FT. On the other hand, the width of $\tilde{g}(x)$ will be $\sim 0.5$ in $x$ if we suppose $\tilde{H}^{Ra}_{0}(z,P_z)=0$ for all the $zP_z>$3.  We conclude the FT of our present results of $\tilde{H}^{Ra}_{0}(z,P_z)$ cannot provide any meaningful constraint on the gluon PDF $g(x)$.

\textit{Summary and outlook: }In summary, we present the first gluon quasi-PDF result for the nucleon and pion with multiple hadron boost momenta $P_z$ and explore different choices of the operators. With proper renormalization, the quasi-PDF matrix elements we obtain agree with the FT of the global-fit PDF within statistical uncertainty, up to mixing from the quark PDF, perturbative matching and higher-twist correction ${\cal O}(1/P_z^2)$. 

Since global fitting results shows that most of the contribution of $g(x)$ comes from the $x<0.1$ region, the width of its FT, $H(zP_z)$, is pretty large as the $H(zP_z)$ becomes half of of its maximum value (at $zP_z$=0) at $zP_z\sim 7$. At the same time, the signal of the lattice simulation and also the validity of the factorization limit us to the small $z$ region. Thus to discern the width of gluon PDF, the lattice simulation with much larger nucleon momentum $P_z$, such as 2-3~GeV, is needed.  To archive a good signal with such a large $P_z$, the momentum smearing~\cite{Bali:2016lva} and cluster decomposition error reduction~\cite{Liu:2017man} should be helpful.

In the theoretical side, the gluon quasi-PDF operator can be renormalized non-perturbatively in the RI/MOM scheme {(the ${\cal O}(F^{z}_{\ \mu}, F^{\mu z};z)$ and ${\cal O}(F^{\mu}_{\ \nu}, F^{\nu \mu};z)$ ($\mu, \nu\neq z$) structures in ${\cal O}_0$ and ${\cal O}_2$ should be renormalized separately before combined together, while ${\cal O}_1$ is multiplicative renormalizable~\cite{Zhang:2018diq,Li:2018tpe})} based on the NPR strategy introduced in Ref.~\cite{Yang:2018bft}, and the matching to the gluon PDF can be calculated perturbatively following the framework used in the quark case~\cite{Stewart:2017tvs}.

\section*{Acknowledgments}

We thank J.W.~Chen, X.D.~Ji, L.~Jin, W. Wang, and J.H.~Zhang for useful discussions,
and the RBC and UKQCD collaborations for providing
us their DWF gauge configurations. ZF, HL and YY are
supported by the US National Science Foundation under
grant PHY 1653405 ``CAREER: Constraining Parton
Distribution Functions for New-Physics Searches''.
KL is partially supported by DOE grant DE-SC0013065 and DOE TMD topical collaboration.
This research used resources of the National Energy Research Scientific Computing Center, a DOE Office of Science User Facility supported by the Office of Science of the U.S. Department of Energy under Contract No. DE-AC02-05CH11231 
through ALCC and ERCAP;
facilities of the USQCD Collaboration, which are funded by the Office of Science of the U.S. Department of Energy,
and supported in part by Michigan State University through computational resources provided by the Institute for Cyber-Enabled Research. 

\bibliographystyle{apsrev4-1}
\bibliography{lattice}

\begin{thebibliography}{28}%
\makeatletter
\providecommand \@ifxundefined [1]{%
 \@ifx{#1\undefined}
}%
\providecommand \@ifnum [1]{%
 \ifnum #1\expandafter \@firstoftwo
 \else \expandafter \@secondoftwo
 \fi
}%
\providecommand \@ifx [1]{%
 \ifx #1\expandafter \@firstoftwo
 \else \expandafter \@secondoftwo
 \fi
}%
\providecommand \natexlab [1]{#1}%
\providecommand \enquote  [1]{``#1''}%
\providecommand \bibnamefont  [1]{#1}%
\providecommand \bibfnamefont [1]{#1}%
\providecommand \citenamefont [1]{#1}%
\providecommand \href@noop [0]{\@secondoftwo}%
\providecommand \href [0]{\begingroup \@sanitize@url \@href}%
\providecommand \@href[1]{\@@startlink{#1}\@@href}%
\providecommand \@@href[1]{\endgroup#1\@@endlink}%
\providecommand \@sanitize@url [0]{\catcode `\\12\catcode `\$12\catcode
  `\&12\catcode `\#12\catcode `\^12\catcode `\_12\catcode `\%12\relax}%
\providecommand \@@startlink[1]{}%
\providecommand \@@endlink[0]{}%
\providecommand \url  [0]{\begingroup\@sanitize@url \@url }%
\providecommand \@url [1]{\endgroup\@href {#1}{\urlprefix }}%
\providecommand \urlprefix  [0]{URL }%
\providecommand \Eprint [0]{\href }%
\providecommand \doibase [0]{http://dx.doi.org/}%
\providecommand \selectlanguage [0]{\@gobble}%
\providecommand \bibinfo  [0]{\@secondoftwo}%
\providecommand \bibfield  [0]{\@secondoftwo}%
\providecommand \translation [1]{[#1]}%
\providecommand \BibitemOpen [0]{}%
\providecommand \bibitemStop [0]{}%
\providecommand \bibitemNoStop [0]{.\EOS\space}%
\providecommand \EOS [0]{\spacefactor3000\relax}%
\providecommand \BibitemShut  [1]{\csname bibitem#1\endcsname}%
\let\auto@bib@innerbib\@empty
\bibitem [{\citenamefont {Czakon}\ \emph {et~al.}(2013)\citenamefont {Czakon},
  \citenamefont {Mangano}, \citenamefont {Mitov},\ and\ \citenamefont
  {Rojo}}]{Czakon:2013tha}%
  \BibitemOpen
  \bibfield  {author} {\bibinfo {author} {\bibfnamefont {M.}~\bibnamefont
  {Czakon}}, \bibinfo {author} {\bibfnamefont {M.~L.}\ \bibnamefont {Mangano}},
  \bibinfo {author} {\bibfnamefont {A.}~\bibnamefont {Mitov}}, \ and\ \bibinfo
  {author} {\bibfnamefont {J.}~\bibnamefont {Rojo}},\ }\href {\doibase
  10.1007/JHEP07(2013)167} {\bibfield  {journal} {\bibinfo  {journal} {JHEP}\
  }\textbf {\bibinfo {volume} {07}},\ \bibinfo {pages} {167} (\bibinfo {year}
  {2013})},\ \Eprint {http://arxiv.org/abs/1303.7215} {arXiv:1303.7215
  [hep-ph]} \BibitemShut {NoStop}%
\bibitem [{\citenamefont {Gauld}\ \emph {et~al.}(2015)\citenamefont {Gauld},
  \citenamefont {Rojo}, \citenamefont {Rottoli},\ and\ \citenamefont
  {Talbert}}]{Gauld:2015yia}%
  \BibitemOpen
  \bibfield  {author} {\bibinfo {author} {\bibfnamefont {R.}~\bibnamefont
  {Gauld}}, \bibinfo {author} {\bibfnamefont {J.}~\bibnamefont {Rojo}},
  \bibinfo {author} {\bibfnamefont {L.}~\bibnamefont {Rottoli}}, \ and\
  \bibinfo {author} {\bibfnamefont {J.}~\bibnamefont {Talbert}},\ }\href
  {\doibase 10.1007/JHEP11(2015)009} {\bibfield  {journal} {\bibinfo  {journal}
  {JHEP}\ }\textbf {\bibinfo {volume} {11}},\ \bibinfo {pages} {009} (\bibinfo
  {year} {2015})},\ \Eprint {http://arxiv.org/abs/1506.08025} {arXiv:1506.08025
  [hep-ph]} \BibitemShut {NoStop}%
\bibitem [{\citenamefont {Horsley}\ \emph {et~al.}(2012)\citenamefont
  {Horsley}, \citenamefont {Millo}, \citenamefont {Nakamura}, \citenamefont
  {Perlt}, \citenamefont {Pleiter}, \citenamefont {Rakow}, \citenamefont
  {Schierholz}, \citenamefont {Schiller}, \citenamefont {Winter},\ and\
  \citenamefont {Zanotti}}]{Horsley:2012pz}%
  \BibitemOpen
  \bibfield  {author} {\bibinfo {author} {\bibfnamefont {R.}~\bibnamefont
  {Horsley}}, \bibinfo {author} {\bibfnamefont {R.}~\bibnamefont {Millo}},
  \bibinfo {author} {\bibfnamefont {Y.}~\bibnamefont {Nakamura}}, \bibinfo
  {author} {\bibfnamefont {H.}~\bibnamefont {Perlt}}, \bibinfo {author}
  {\bibfnamefont {D.}~\bibnamefont {Pleiter}}, \bibinfo {author} {\bibfnamefont
  {P.~E.~L.}\ \bibnamefont {Rakow}}, \bibinfo {author} {\bibfnamefont
  {G.}~\bibnamefont {Schierholz}}, \bibinfo {author} {\bibfnamefont
  {A.}~\bibnamefont {Schiller}}, \bibinfo {author} {\bibfnamefont
  {F.}~\bibnamefont {Winter}}, \ and\ \bibinfo {author} {\bibfnamefont {J.~M.}\
  \bibnamefont {Zanotti}} (\bibinfo {collaboration} {UKQCD, QCDSF}),\ }\href
  {\doibase 10.1016/j.physletb.2012.07.004} {\bibfield  {journal} {\bibinfo
  {journal} {Phys. Lett.}\ }\textbf {\bibinfo {volume} {B714}},\ \bibinfo
  {pages} {312} (\bibinfo {year} {2012})},\ \Eprint
  {http://arxiv.org/abs/1205.6410} {arXiv:1205.6410 [hep-lat]} \BibitemShut
  {NoStop}%
\bibitem [{\citenamefont {Deka}\ \emph {et~al.}(2015)\citenamefont {Deka} \emph
  {et~al.}}]{Deka:2013zha}%
  \BibitemOpen
  \bibfield  {author} {\bibinfo {author} {\bibfnamefont {M.}~\bibnamefont
  {Deka}} \emph {et~al.},\ }\href {\doibase 10.1103/PhysRevD.91.014505}
  {\bibfield  {journal} {\bibinfo  {journal} {Phys. Rev.}\ }\textbf {\bibinfo
  {volume} {D91}},\ \bibinfo {pages} {014505} (\bibinfo {year} {2015})},\
  \Eprint {http://arxiv.org/abs/1312.4816} {arXiv:1312.4816 [hep-lat]}
  \BibitemShut {NoStop}%
\bibitem [{\citenamefont {Alexandrou}\ \emph
  {et~al.}(2017{\natexlab{a}})\citenamefont {Alexandrou}, \citenamefont
  {Constantinou}, \citenamefont {Hadjiyiannakou}, \citenamefont {Jansen},
  \citenamefont {Panagopoulos},\ and\ \citenamefont
  {Wiese}}]{Alexandrou:2016ekb}%
  \BibitemOpen
  \bibfield  {author} {\bibinfo {author} {\bibfnamefont {C.}~\bibnamefont
  {Alexandrou}}, \bibinfo {author} {\bibfnamefont {M.}~\bibnamefont
  {Constantinou}}, \bibinfo {author} {\bibfnamefont {K.}~\bibnamefont
  {Hadjiyiannakou}}, \bibinfo {author} {\bibfnamefont {K.}~\bibnamefont
  {Jansen}}, \bibinfo {author} {\bibfnamefont {H.}~\bibnamefont
  {Panagopoulos}}, \ and\ \bibinfo {author} {\bibfnamefont {C.}~\bibnamefont
  {Wiese}},\ }\href {\doibase 10.1103/PhysRevD.96.054503} {\bibfield  {journal}
  {\bibinfo  {journal} {Phys. Rev.}\ }\textbf {\bibinfo {volume} {D96}},\
  \bibinfo {pages} {054503} (\bibinfo {year} {2017}{\natexlab{a}})},\ \Eprint
  {http://arxiv.org/abs/1611.06901} {arXiv:1611.06901 [hep-lat]} \BibitemShut
  {NoStop}%
\bibitem [{\citenamefont {Alexandrou}\ \emph
  {et~al.}(2017{\natexlab{b}})\citenamefont {Alexandrou}, \citenamefont
  {Constantinou}, \citenamefont {Hadjiyiannakou}, \citenamefont {Jansen},
  \citenamefont {Kallidonis}, \citenamefont {Koutsou}, \citenamefont {Vaquero
  Avil{\'e}s-Casco},\ and\ \citenamefont {Wiese}}]{Alexandrou:2017oeh}%
  \BibitemOpen
  \bibfield  {author} {\bibinfo {author} {\bibfnamefont {C.}~\bibnamefont
  {Alexandrou}}, \bibinfo {author} {\bibfnamefont {M.}~\bibnamefont
  {Constantinou}}, \bibinfo {author} {\bibfnamefont {K.}~\bibnamefont
  {Hadjiyiannakou}}, \bibinfo {author} {\bibfnamefont {K.}~\bibnamefont
  {Jansen}}, \bibinfo {author} {\bibfnamefont {C.}~\bibnamefont {Kallidonis}},
  \bibinfo {author} {\bibfnamefont {G.}~\bibnamefont {Koutsou}}, \bibinfo
  {author} {\bibfnamefont {A.}~\bibnamefont {Vaquero Avil{\'e}s-Casco}}, \ and\
  \bibinfo {author} {\bibfnamefont {C.}~\bibnamefont {Wiese}},\ }\href
  {\doibase 10.1103/PhysRevLett.119.142002} {\bibfield  {journal} {\bibinfo
  {journal} {Phys. Rev. Lett.}\ }\textbf {\bibinfo {volume} {119}},\ \bibinfo
  {pages} {142002} (\bibinfo {year} {2017}{\natexlab{b}})},\ \Eprint
  {http://arxiv.org/abs/1706.02973} {arXiv:1706.02973 [hep-lat]} \BibitemShut
  {NoStop}%
\bibitem [{\citenamefont {Yang}\ \emph
  {et~al.}(2018{\natexlab{a}})\citenamefont {Yang}, \citenamefont {Gong},
  \citenamefont {Liang}, \citenamefont {Lin}, \citenamefont {Liu},
  \citenamefont {Pefkou},\ and\ \citenamefont {Shanahan}}]{Yang:2018bft}%
  \BibitemOpen
  \bibfield  {author} {\bibinfo {author} {\bibfnamefont {Y.-B.}\ \bibnamefont
  {Yang}}, \bibinfo {author} {\bibfnamefont {M.}~\bibnamefont {Gong}}, \bibinfo
  {author} {\bibfnamefont {J.}~\bibnamefont {Liang}}, \bibinfo {author}
  {\bibfnamefont {H.-W.}\ \bibnamefont {Lin}}, \bibinfo {author} {\bibfnamefont
  {K.-F.}\ \bibnamefont {Liu}}, \bibinfo {author} {\bibfnamefont
  {D.}~\bibnamefont {Pefkou}}, \ and\ \bibinfo {author} {\bibfnamefont
  {P.}~\bibnamefont {Shanahan}},\ }\href@noop {} {\  (\bibinfo {year}
  {2018}{\natexlab{a}})},\ \Eprint {http://arxiv.org/abs/1805.00531}
  {arXiv:1805.00531 [hep-lat]} \BibitemShut {NoStop}%
\bibitem [{\citenamefont {Yang}\ \emph
  {et~al.}(2018{\natexlab{b}})\citenamefont {Yang}, \citenamefont {Liang},
  \citenamefont {Bi}, \citenamefont {Chen}, \citenamefont {Draper},
  \citenamefont {Liu},\ and\ \citenamefont {Liu}}]{Yang:2018nqn}%
  \BibitemOpen
  \bibfield  {author} {\bibinfo {author} {\bibfnamefont {Y.-B.}\ \bibnamefont
  {Yang}}, \bibinfo {author} {\bibfnamefont {J.}~\bibnamefont {Liang}},
  \bibinfo {author} {\bibfnamefont {Y.-J.}\ \bibnamefont {Bi}}, \bibinfo
  {author} {\bibfnamefont {Y.}~\bibnamefont {Chen}}, \bibinfo {author}
  {\bibfnamefont {T.}~\bibnamefont {Draper}}, \bibinfo {author} {\bibfnamefont
  {K.-F.}\ \bibnamefont {Liu}}, \ and\ \bibinfo {author} {\bibfnamefont
  {Z.}~\bibnamefont {Liu}},\ }\href@noop {} {\  (\bibinfo {year}
  {2018}{\natexlab{b}})},\ \Eprint {http://arxiv.org/abs/1808.08677}
  {arXiv:1808.08677 [hep-lat]} \BibitemShut {NoStop}%
\bibitem [{\citenamefont {Ji}(2013)}]{Ji:2013dva}%
  \BibitemOpen
  \bibfield  {author} {\bibinfo {author} {\bibfnamefont {X.}~\bibnamefont
  {Ji}},\ }\href {\doibase 10.1103/PhysRevLett.110.262002} {\bibfield
  {journal} {\bibinfo  {journal} {Phys. Rev. Lett.}\ }\textbf {\bibinfo
  {volume} {110}},\ \bibinfo {pages} {262002} (\bibinfo {year} {2013})},\
  \Eprint {http://arxiv.org/abs/1305.1539} {arXiv:1305.1539 [hep-ph]}
  \BibitemShut {NoStop}%
\bibitem [{\citenamefont {Ji}(2014)}]{Ji:2014gla}%
  \BibitemOpen
  \bibfield  {author} {\bibinfo {author} {\bibfnamefont {X.}~\bibnamefont
  {Ji}},\ }\href {\doibase 10.1007/s11433-014-5492-3} {\bibfield  {journal}
  {\bibinfo  {journal} {Sci. China Phys. Mech. Astron.}\ }\textbf {\bibinfo
  {volume} {57}},\ \bibinfo {pages} {1407} (\bibinfo {year} {2014})},\ \Eprint
  {http://arxiv.org/abs/1404.6680} {arXiv:1404.6680 [hep-ph]} \BibitemShut
  {NoStop}%
\bibitem [{\citenamefont {Zhang}\ \emph {et~al.}(2018)\citenamefont {Zhang},
  \citenamefont {Ji}, \citenamefont {Sch{\"a}fer}, \citenamefont {Wang},\ and\
  \citenamefont {Zhao}}]{Zhang:2018diq}%
  \BibitemOpen
  \bibfield  {author} {\bibinfo {author} {\bibfnamefont {J.-H.}\ \bibnamefont
  {Zhang}}, \bibinfo {author} {\bibfnamefont {X.}~\bibnamefont {Ji}}, \bibinfo
  {author} {\bibfnamefont {A.}~\bibnamefont {Sch{\"a}fer}}, \bibinfo {author}
  {\bibfnamefont {W.}~\bibnamefont {Wang}}, \ and\ \bibinfo {author}
  {\bibfnamefont {S.}~\bibnamefont {Zhao}},\ }\href@noop {} {\  (\bibinfo
  {year} {2018})},\ \Eprint {http://arxiv.org/abs/1808.10824} {arXiv:1808.10824
  [hep-ph]} \BibitemShut {NoStop}%
\bibitem [{\citenamefont {Li}\ \emph {et~al.}(2018)\citenamefont {Li},
  \citenamefont {Ma},\ and\ \citenamefont {Qiu}}]{Li:2018tpe}%
  \BibitemOpen
  \bibfield  {author} {\bibinfo {author} {\bibfnamefont {Z.-Y.}\ \bibnamefont
  {Li}}, \bibinfo {author} {\bibfnamefont {Y.-Q.}\ \bibnamefont {Ma}}, \ and\
  \bibinfo {author} {\bibfnamefont {J.-W.}\ \bibnamefont {Qiu}},\ }\href@noop
  {} {\  (\bibinfo {year} {2018})},\ \Eprint {http://arxiv.org/abs/1809.01836}
  {arXiv:1809.01836 [hep-ph]} \BibitemShut {NoStop}%
\bibitem [{\citenamefont {Wang}\ \emph {et~al.}(2018)\citenamefont {Wang},
  \citenamefont {Zhao},\ and\ \citenamefont {Zhu}}]{Wang:2017qyg}%
  \BibitemOpen
  \bibfield  {author} {\bibinfo {author} {\bibfnamefont {W.}~\bibnamefont
  {Wang}}, \bibinfo {author} {\bibfnamefont {S.}~\bibnamefont {Zhao}}, \ and\
  \bibinfo {author} {\bibfnamefont {R.}~\bibnamefont {Zhu}},\ }\href {\doibase
  10.1140/epjc/s10052-018-5617-3} {\bibfield  {journal} {\bibinfo  {journal}
  {Eur. Phys. J.}\ }\textbf {\bibinfo {volume} {C78}},\ \bibinfo {pages} {147}
  (\bibinfo {year} {2018})},\ \Eprint {http://arxiv.org/abs/1708.02458}
  {arXiv:1708.02458 [hep-ph]} \BibitemShut {NoStop}%
\bibitem [{\citenamefont {Wang}\ and\ \citenamefont
  {Zhao}(2018)}]{Wang:2017eel}%
  \BibitemOpen
  \bibfield  {author} {\bibinfo {author} {\bibfnamefont {W.}~\bibnamefont
  {Wang}}\ and\ \bibinfo {author} {\bibfnamefont {S.}~\bibnamefont {Zhao}},\
  }\href {\doibase 10.1007/JHEP05(2018)142} {\bibfield  {journal} {\bibinfo
  {journal} {JHEP}\ }\textbf {\bibinfo {volume} {05}},\ \bibinfo {pages} {142}
  (\bibinfo {year} {2018})},\ \Eprint {http://arxiv.org/abs/1712.09247}
  {arXiv:1712.09247 [hep-ph]} \BibitemShut {NoStop}%
\bibitem [{\citenamefont {Liu}\ \emph {et~al.}(2018{\natexlab{a}})\citenamefont
  {Liu}, \citenamefont {Chen}, \citenamefont {Jin}, \citenamefont {Lin},
  \citenamefont {Yang}, \citenamefont {Zhang},\ and\ \citenamefont
  {Zhao}}]{Liu:2018uuj}%
  \BibitemOpen
  \bibfield  {author} {\bibinfo {author} {\bibfnamefont {Y.-S.}\ \bibnamefont
  {Liu}}, \bibinfo {author} {\bibfnamefont {J.-W.}\ \bibnamefont {Chen}},
  \bibinfo {author} {\bibfnamefont {L.}~\bibnamefont {Jin}}, \bibinfo {author}
  {\bibfnamefont {H.-W.}\ \bibnamefont {Lin}}, \bibinfo {author} {\bibfnamefont
  {Y.-B.}\ \bibnamefont {Yang}}, \bibinfo {author} {\bibfnamefont {J.-H.}\
  \bibnamefont {Zhang}}, \ and\ \bibinfo {author} {\bibfnamefont
  {Y.}~\bibnamefont {Zhao}},\ }\href@noop {} {\  (\bibinfo {year}
  {2018}{\natexlab{a}})},\ \Eprint {http://arxiv.org/abs/1807.06566}
  {arXiv:1807.06566 [hep-lat]} \BibitemShut {NoStop}%
\bibitem [{\citenamefont {Radyushkin}(2017)}]{Radyushkin:2017cyf}%
  \BibitemOpen
  \bibfield  {author} {\bibinfo {author} {\bibfnamefont {A.~V.}\ \bibnamefont
  {Radyushkin}},\ }\href {\doibase 10.1103/PhysRevD.96.034025} {\bibfield
  {journal} {\bibinfo  {journal} {Phys. Rev.}\ }\textbf {\bibinfo {volume}
  {D96}},\ \bibinfo {pages} {034025} (\bibinfo {year} {2017})},\ \Eprint
  {http://arxiv.org/abs/1705.01488} {arXiv:1705.01488 [hep-ph]} \BibitemShut
  {NoStop}%
\bibitem [{\citenamefont {Orginos}\ \emph {et~al.}(2017)\citenamefont
  {Orginos}, \citenamefont {Radyushkin}, \citenamefont {Karpie},\ and\
  \citenamefont {Zafeiropoulos}}]{Orginos:2017kos}%
  \BibitemOpen
  \bibfield  {author} {\bibinfo {author} {\bibfnamefont {K.}~\bibnamefont
  {Orginos}}, \bibinfo {author} {\bibfnamefont {A.}~\bibnamefont {Radyushkin}},
  \bibinfo {author} {\bibfnamefont {J.}~\bibnamefont {Karpie}}, \ and\ \bibinfo
  {author} {\bibfnamefont {S.}~\bibnamefont {Zafeiropoulos}},\ }\href {\doibase
  10.1103/PhysRevD.96.094503} {\bibfield  {journal} {\bibinfo  {journal} {Phys.
  Rev.}\ }\textbf {\bibinfo {volume} {D96}},\ \bibinfo {pages} {094503}
  (\bibinfo {year} {2017})},\ \Eprint {http://arxiv.org/abs/1706.05373}
  {arXiv:1706.05373 [hep-ph]} \BibitemShut {NoStop}%
\bibitem [{\citenamefont {Broniowski}\ and\ \citenamefont
  {Ruiz~Arriola}(2018)}]{Broniowski:2017gfp}%
  \BibitemOpen
  \bibfield  {author} {\bibinfo {author} {\bibfnamefont {W.}~\bibnamefont
  {Broniowski}}\ and\ \bibinfo {author} {\bibfnamefont {E.}~\bibnamefont
  {Ruiz~Arriola}},\ }\href {\doibase 10.1103/PhysRevD.97.034031} {\bibfield
  {journal} {\bibinfo  {journal} {Phys. Rev.}\ }\textbf {\bibinfo {volume}
  {D97}},\ \bibinfo {pages} {034031} (\bibinfo {year} {2018})},\ \Eprint
  {http://arxiv.org/abs/1711.03377} {arXiv:1711.03377 [hep-ph]} \BibitemShut
  {NoStop}%
\bibitem [{\citenamefont {Aoki}\ \emph {et~al.}(2011)\citenamefont {Aoki} \emph
  {et~al.}}]{Aoki:2010dy}%
  \BibitemOpen
  \bibfield  {author} {\bibinfo {author} {\bibfnamefont {Y.}~\bibnamefont
  {Aoki}} \emph {et~al.} (\bibinfo {collaboration} {RBC, UKQCD}),\ }\href
  {\doibase 10.1103/PhysRevD.83.074508} {\bibfield  {journal} {\bibinfo
  {journal} {Phys. Rev.}\ }\textbf {\bibinfo {volume} {D83}},\ \bibinfo {pages}
  {074508} (\bibinfo {year} {2011})},\ \Eprint {http://arxiv.org/abs/1011.0892}
  {arXiv:1011.0892 [hep-lat]} \BibitemShut {NoStop}%
\bibitem [{\citenamefont {Li}\ \emph {et~al.}(2010)\citenamefont {Li},
  \citenamefont {Alexandru}, \citenamefont {Chen}, \citenamefont {Doi},
  \citenamefont {Dong}, \citenamefont {Draper}, \citenamefont {Gong},
  \citenamefont {Hasenfratz}, \citenamefont {Horvath}, \citenamefont {Lee}
  \emph {et~al.}}]{grid1}%
  \BibitemOpen
  \bibfield  {author} {\bibinfo {author} {\bibfnamefont {A.}~\bibnamefont
  {Li}}, \bibinfo {author} {\bibfnamefont {A.}~\bibnamefont {Alexandru}},
  \bibinfo {author} {\bibfnamefont {Y.}~\bibnamefont {Chen}}, \bibinfo {author}
  {\bibfnamefont {T.}~\bibnamefont {Doi}}, \bibinfo {author} {\bibfnamefont
  {S.}~\bibnamefont {Dong}}, \bibinfo {author} {\bibfnamefont {T.}~\bibnamefont
  {Draper}}, \bibinfo {author} {\bibfnamefont {M.}~\bibnamefont {Gong}},
  \bibinfo {author} {\bibfnamefont {A.}~\bibnamefont {Hasenfratz}}, \bibinfo
  {author} {\bibfnamefont {I.}~\bibnamefont {Horvath}}, \bibinfo {author}
  {\bibfnamefont {F.}~\bibnamefont {Lee}},  \emph {et~al.},\ }\href@noop {}
  {\bibfield  {journal} {\bibinfo  {journal} {Physical Review D}\ }\textbf
  {\bibinfo {volume} {82}},\ \bibinfo {pages} {114501} (\bibinfo {year}
  {2010})}\BibitemShut {NoStop}%
\bibitem [{\citenamefont {Gong}\ \emph {et~al.}(2013)\citenamefont {Gong},
  \citenamefont {Alexandru}, \citenamefont {Chen}, \citenamefont {Doi},
  \citenamefont {Dong}, \citenamefont {Draper}, \citenamefont {Freeman},
  \citenamefont {Glatzmaier}, \citenamefont {Li}, \citenamefont {Liu} \emph
  {et~al.}}]{grid2}%
  \BibitemOpen
  \bibfield  {author} {\bibinfo {author} {\bibfnamefont {M.}~\bibnamefont
  {Gong}}, \bibinfo {author} {\bibfnamefont {A.}~\bibnamefont {Alexandru}},
  \bibinfo {author} {\bibfnamefont {Y.}~\bibnamefont {Chen}}, \bibinfo {author}
  {\bibfnamefont {T.}~\bibnamefont {Doi}}, \bibinfo {author} {\bibfnamefont
  {S.}~\bibnamefont {Dong}}, \bibinfo {author} {\bibfnamefont {T.}~\bibnamefont
  {Draper}}, \bibinfo {author} {\bibfnamefont {W.}~\bibnamefont {Freeman}},
  \bibinfo {author} {\bibfnamefont {M.}~\bibnamefont {Glatzmaier}}, \bibinfo
  {author} {\bibfnamefont {A.}~\bibnamefont {Li}}, \bibinfo {author}
  {\bibfnamefont {K.}~\bibnamefont {Liu}},  \emph {et~al.},\ }\href@noop {}
  {\bibfield  {journal} {\bibinfo  {journal} {Physical Review D}\ }\textbf
  {\bibinfo {volume} {88}},\ \bibinfo {pages} {014503} (\bibinfo {year}
  {2013})}\BibitemShut {NoStop}%
\bibitem [{\citenamefont {Berkowitz}\ \emph {et~al.}(2017)\citenamefont
  {Berkowitz} \emph {et~al.}}]{Berkowitz:2017gql}%
  \BibitemOpen
  \bibfield  {author} {\bibinfo {author} {\bibfnamefont {E.}~\bibnamefont
  {Berkowitz}} \emph {et~al.},\ }\href@noop {} {\  (\bibinfo {year} {2017})},\
  \Eprint {http://arxiv.org/abs/1704.01114} {arXiv:1704.01114 [hep-lat]}
  \BibitemShut {NoStop}%
\bibitem [{\citenamefont {Chang}\ \emph {et~al.}(2018)\citenamefont {Chang}
  \emph {et~al.}}]{Chang:2018uxx}%
  \BibitemOpen
  \bibfield  {author} {\bibinfo {author} {\bibfnamefont {C.~C.}\ \bibnamefont
  {Chang}} \emph {et~al.},\ }\href {\doibase 10.1038/s41586-018-0161-8}
  {\bibfield  {journal} {\bibinfo  {journal} {Nature}\ }\textbf {\bibinfo
  {volume} {558}},\ \bibinfo {pages} {91} (\bibinfo {year} {2018})},\ \Eprint
  {http://arxiv.org/abs/1805.12130} {arXiv:1805.12130 [hep-lat]} \BibitemShut
  {NoStop}%
\bibitem [{\citenamefont {Dulat}\ \emph {et~al.}(2016)\citenamefont {Dulat},
  \citenamefont {Hou}, \citenamefont {Gao}, \citenamefont {Guzzi},
  \citenamefont {Huston}, \citenamefont {Nadolsky}, \citenamefont {Pumplin},
  \citenamefont {Schmidt}, \citenamefont {Stump},\ and\ \citenamefont
  {Yuan}}]{Dulat:2015mca}%
  \BibitemOpen
  \bibfield  {author} {\bibinfo {author} {\bibfnamefont {S.}~\bibnamefont
  {Dulat}}, \bibinfo {author} {\bibfnamefont {T.-J.}\ \bibnamefont {Hou}},
  \bibinfo {author} {\bibfnamefont {J.}~\bibnamefont {Gao}}, \bibinfo {author}
  {\bibfnamefont {M.}~\bibnamefont {Guzzi}}, \bibinfo {author} {\bibfnamefont
  {J.}~\bibnamefont {Huston}}, \bibinfo {author} {\bibfnamefont
  {P.}~\bibnamefont {Nadolsky}}, \bibinfo {author} {\bibfnamefont
  {J.}~\bibnamefont {Pumplin}}, \bibinfo {author} {\bibfnamefont
  {C.}~\bibnamefont {Schmidt}}, \bibinfo {author} {\bibfnamefont
  {D.}~\bibnamefont {Stump}}, \ and\ \bibinfo {author} {\bibfnamefont {C.~P.}\
  \bibnamefont {Yuan}},\ }\href {\doibase 10.1103/PhysRevD.93.033006}
  {\bibfield  {journal} {\bibinfo  {journal} {Phys. Rev.}\ }\textbf {\bibinfo
  {volume} {D93}},\ \bibinfo {pages} {033006} (\bibinfo {year} {2016})},\
  \Eprint {http://arxiv.org/abs/1506.07443} {arXiv:1506.07443 [hep-ph]}
  \BibitemShut {NoStop}%
\bibitem [{\citenamefont {Butterworth}\ \emph {et~al.}(2016)\citenamefont
  {Butterworth} \emph {et~al.}}]{Butterworth:2015oua}%
  \BibitemOpen
  \bibfield  {author} {\bibinfo {author} {\bibfnamefont {J.}~\bibnamefont
  {Butterworth}} \emph {et~al.},\ }\href {\doibase
  10.1088/0954-3899/43/2/023001} {\bibfield  {journal} {\bibinfo  {journal} {J.
  Phys.}\ }\textbf {\bibinfo {volume} {G43}},\ \bibinfo {pages} {023001}
  (\bibinfo {year} {2016})},\ \Eprint {http://arxiv.org/abs/1510.03865}
  {arXiv:1510.03865 [hep-ph]} \BibitemShut {NoStop}%
\bibitem [{\citenamefont {Bali}\ \emph {et~al.}(2016)\citenamefont {Bali},
  \citenamefont {Lang}, \citenamefont {Musch},\ and\ \citenamefont
  {Sch{\"a}fer}}]{Bali:2016lva}%
  \BibitemOpen
  \bibfield  {author} {\bibinfo {author} {\bibfnamefont {G.~S.}\ \bibnamefont
  {Bali}}, \bibinfo {author} {\bibfnamefont {B.}~\bibnamefont {Lang}}, \bibinfo
  {author} {\bibfnamefont {B.~U.}\ \bibnamefont {Musch}}, \ and\ \bibinfo
  {author} {\bibfnamefont {A.}~\bibnamefont {Sch{\"a}fer}},\ }\href {\doibase
  10.1103/PhysRevD.93.094515} {\bibfield  {journal} {\bibinfo  {journal} {Phys.
  Rev.}\ }\textbf {\bibinfo {volume} {D93}},\ \bibinfo {pages} {094515}
  (\bibinfo {year} {2016})},\ \Eprint {http://arxiv.org/abs/1602.05525}
  {arXiv:1602.05525 [hep-lat]} \BibitemShut {NoStop}%
\bibitem [{\citenamefont {Liu}\ \emph {et~al.}(2018{\natexlab{b}})\citenamefont
  {Liu}, \citenamefont {Liang},\ and\ \citenamefont {Yang}}]{Liu:2017man}%
  \BibitemOpen
  \bibfield  {author} {\bibinfo {author} {\bibfnamefont {K.-F.}\ \bibnamefont
  {Liu}}, \bibinfo {author} {\bibfnamefont {J.}~\bibnamefont {Liang}}, \ and\
  \bibinfo {author} {\bibfnamefont {Y.-B.}\ \bibnamefont {Yang}},\ }\href
  {\doibase 10.1103/PhysRevD.97.034507} {\bibfield  {journal} {\bibinfo
  {journal} {Phys. Rev.}\ }\textbf {\bibinfo {volume} {D97}},\ \bibinfo {pages}
  {034507} (\bibinfo {year} {2018}{\natexlab{b}})},\ \Eprint
  {http://arxiv.org/abs/1705.06358} {arXiv:1705.06358 [hep-lat]} \BibitemShut
  {NoStop}%
\bibitem [{\citenamefont {Stewart}\ and\ \citenamefont
  {Zhao}(2018)}]{Stewart:2017tvs}%
  \BibitemOpen
  \bibfield  {author} {\bibinfo {author} {\bibfnamefont {I.~W.}\ \bibnamefont
  {Stewart}}\ and\ \bibinfo {author} {\bibfnamefont {Y.}~\bibnamefont {Zhao}},\
  }\href {\doibase 10.1103/PhysRevD.97.054512} {\bibfield  {journal} {\bibinfo
  {journal} {Phys. Rev.}\ }\textbf {\bibinfo {volume} {D97}},\ \bibinfo {pages}
  {054512} (\bibinfo {year} {2018})},\ \Eprint
  {http://arxiv.org/abs/1709.04933} {arXiv:1709.04933 [hep-ph]} \BibitemShut
  {NoStop}%
\end{thebibliography}%

\end{document}